\documentclass[a4paper]{article}
\usepackage{setspace}
\usepackage[margin=1in]{geometry}
\usepackage{amsmath, amssymb, graphicx, tabularx, float}
\usepackage[sectionbib]{natbib}
\usepackage[para]{threeparttable}

\providecommand{\keywords}[1]
{
  \small	
  \textbf{\textit{Keywords:}} #1
}

\begin{document}

\title{Penalized estimation for single-index varying-coefficient models
with applications to integrative genomic analysis}

\author{Hoi Min Ng, Binyan Jiang, Kin Yau Wong\thanks{Corresponding author; email: kin-yau.wong@polyu.edu.hk}\\[4pt]The Hong Kong Polytechnic University}

\date{}

\maketitle

\begin{abstract}
Recent technological advances have made it possible to collect high-dimensional genomic data along with clinical data on a large number of subjects. In the studies of chronic diseases such as cancer, it is of great interest to integrate clinical and genomic data to build a comprehensive understanding of the disease mechanisms. Despite extensive studies on integrative analysis, it remains an ongoing challenge to model the interaction effects between clinical and genomic variables, due to high-dimensionality of the data and heterogeneity across data types. In this paper, we propose an integrative approach that models interaction effects using a single-index varying-coefficient model, where the effects of genomic features can be modified by clinical variables. We propose a penalized approach for separate selection of main and interaction effects. We demonstrate the advantages of the proposed methods through extensive simulation studies and provide applications to a motivating cancer genomic study.
\end{abstract}


\section{Introduction}
The major goals of cancer genomics include identification of risk factors associated with the progression of cancer and prediction of disease outcomes. In conventional cancer studies, clinical factors such as age, gender, and tumor stage are routinely studied and used as prognostic factors. Recent advances in high-throughput technologies facilitate the generation of high-dimensional genomic data, which provide useful insights into the molecular pathways underlying cancer development. For example, in The Cancer Genome Atlas (TCGA), clinical and omics data, including copy number alteration, DNA methylation, mutation, and the expressions of mRNA, microRNA, and protein, were collected from more than 11 000 cancer patients across 33 tumor types. Also, in the Molecular Taxonomy of Breast Cancer International Consortium \citep{curtis2012genomic}, copy number alteration, mutation, and mRNA expression data were collected from about 2 000 breast cancer patients. Such massive omics data enable researchers to gain deeper understanding of the biological mechanisms involved in cancer progression. Many studies have shown that the integrative analysis of clinical and genomic data confers greater prognostic power than the analysis of clinical data alone \citep{li2006survival, shedden2008gene, bovelstad2009survival, fan2011building, zhao2015combining}.

Methods for integrative analysis of clinical and genomic data have been extensively investigated in recent decades. A straight-forward integration strategy is to combine clinical and genomic data into a single data set, on which conventional analyses are performed. Some studies demonstrated that direct combination of clinical factors and gene expressions improves risk prediction over the use of either data type alone \citep{bovelstad2009survival, fan2011building, zhao2015combining}. Alternatively, one may take into account the difference in prognostic power of the data types through some weighting approach. \cite{gevaert2006predicting} developed a Bayesian network approach that builds separate models for clinical and microarray data and used a weighted approach to combine the model predictions. \cite{daemen2007integration} proposed a weighted kernel-based method to integrate clinical and microarray data for classification. Both studies demonstrated that models that account for the distinction between clinical and genomic data tend to yield better prediction accuracy over models that treat these data types equally. However, these methods do not consider
interaction effects between genomic
and clinical variables, that is,
modifications of the effects of genomic variables by clinical variables.

Integrative methods for multiple genomic data types have also been studied.
\cite{lanckriet2004statistical}, \cite{daemen2009kernel}, and \cite{seoane2014pathway} proposed weighted kernel-based approaches to integrate multiple heterogeneous data types. \cite{boulesteix2017ipf} and \cite{wong2019boost} proposed penalization regression methods on multiple data types while accounting for their differences in prognostic power. \cite{wang2013ibag} and \cite{zhu2016integrating}
incorporated prior knowledge of regulatory relationship
among different types of genomic variables for the regression of
disease outcomes on genomic variables.
%incorporated the association structure among different data types analyzed the regulation of gene expressions by epigenetic or genetic features and linked the regulators with patient survival in a regression model.
These methods, though accounting for differences among different data types, do not allow for interaction effects. \cite{nevins2003towards} and \cite{pittman2004integrated} developed tree-based classification methods to evaluate the effects of clinical and genomic data on (binary) disease outcomes, allowing for potential interactions among multiple risk factors. However, the estimated model does not have simple interpretations, and the methods may not accommodate a large number of variables. In a recent study, \cite{li2020semiparametric} proposed a regularization method to select for gene-gene interaction effects on disease outcomes,
but the interactions between clinical and genomic data were not considered.

The effects of genomic features on cancer progression are often modified by clinical factors. For example, \cite{landi2008gene} demonstrated that the effects of some gene expressions on the risk of lung cancer mortality vary with tobacco consumption. Also, \cite{chen2017differentiated} and \cite{relli2018distinct} showed that the molecular mechanisms of carcinogenesis exhibit a high level of heterogeneity between two subtypes of non-small-cell lung carcinoma (NSCLC), and the same set of features can have distinct effects on disease outcome across different subtypes. As the effects of genomic features can vary across different clinical characteristics, it is highly desirable to incorporate interaction effects between clinical and genomic variables in regression analyses of disease outcomes on clinical and genomic variables.

A conventional approach to model interaction effects is to include pairwise product terms of predictors into the regression model. However, this approach may not be suitable  for analyzing the interactions between clinical and genomic data. First, adding product terms may greatly expand the model complexity and aggravate the high-dimensionality issue. Second, the scales of (quantitative) clinical and genomic variables are generally incomparable, and modeling interaction effects using pairwise product terms may not be appropriate.
%Furthermore, it is impractical to interpret the interaction effect between two continuous variables.

To address the aforementioned issues, we propose a single-index varying-coefficient model to accommodate potential interaction effects between clinical and genomic features. The single-index varying-coefficient model combines the varying-coefficient model \citep{hastie1993varying} and the single-index model \citep{hardle1993optimal}. It allows the effects of genomic features to vary flexibly with a single index, which is a linear combination of clinical features. This model avoids the curse of dimensionality by projecting the clinical features to an index, so the number of parameters only increases linearly with the number of features.
%This stands in contrast to including all possible pairwise interactions that the number of parameters can be substantially large.
Also, to accommodate the difference in scales between clinical and genomic features, effects of genomic features are formulated
as nonparametric functions of the index.

We propose a penalized (sieve) maximum likelihood estimation method for variable selection and estimation. In particular,
we adopt a novel two-part penalty, which allows for separate
selection of genomic features with effects modified by clinical features and of genomic features with non-zero constant effect.
A coordinate-wise algorithm for the computation of the
penalized estimators is developed.
%is developed to update the single-index parameters and select regression coefficients alternately.
The proposed methods can be applied to common types of outcome variables, including continuous, binary, and censored outcomes. 

%In this paper, we develop a partial linear single-index varying-coefficient model to characterize the main and interaction effects on the outcome of interest. The proposed method allows the effects of (high-dimensional) predictors to be characterized by another set of low-dimensional predictors. We propose a penalized (sieve) maximum likelihood estimation method that effectively identifies important predictors and distinguishes the effects modified by the single-index. A coordinate-wise algorithm is applied to update the single-index parameters and select regression coefficients alternately. Our proposed method accommodates many types of outcome variables, including continuous, binary, and censored outcomes. 

The rest of this paper is organized as follows. We describe the model and estimation procedures in Section \ref{sec:model}. We assess the estimation performance of the proposed methods through simulation studies, and the results are summarized in Section \ref{sec:sim}. We demonstrate the applications of the proposed methods on two TCGA data sets in Section \ref{sec:rda}. Finally, we make some concluding remarks in Section \ref{sec:diss}. Computation details and additional numerical results are given in the Appendix.

\section{Model and estimation}
\label{sec:model}
\subsection{Model, data, and likelihood}
Let $Y$ be an outcome of interest, $\boldsymbol{U}$ and $\boldsymbol{Z}$ be two sets of low-dimensional predictors that may overlap, and $\boldsymbol{X}\equiv(X_{0},\ldots,X_{p})^{\mathrm{T}}$ be a set of potentially high-dimensional predictors with $X_{0}=1$.
We are interested in the effect of $(\boldsymbol{X},\boldsymbol{Z})$ on $Y$, where the effect of $\boldsymbol{X}$ is allowed to depend on $\boldsymbol{U}$. We assume the following partial linear single-index varying-coefficient model:
\begin{align}
Y\mid(\boldsymbol{U},\boldsymbol{X},\boldsymbol{Z})\sim f\bigg\{\cdot\;;\sum_{j=0}^{p}g_{j}(\boldsymbol{U}^{\mathrm{T}}\boldsymbol{\beta})X_{j}+\boldsymbol{Z}^{\mathrm{\mathrm{T}}}\boldsymbol{\psi}\bigg\},\label{eq:model}
\end{align}
where $f$ is a density function, $\boldsymbol{\beta}$ and $\boldsymbol{\psi}$ are regression parameters, and $g_{0},\ldots,g_{p}$ are unspecified smooth functions. For model identifiability, we set $\Vert\boldsymbol{\beta}\Vert=1$, and if $\boldsymbol{U}$ is a subset of $\boldsymbol{Z}$, then we set the component of $\boldsymbol{\psi}$ that corresponds to the last component of $\boldsymbol{U}$ to be 0. This model assumes that the effect of each component of $\boldsymbol{X}$ is characterized by a nonparametric transformation of an index $\boldsymbol{U}^{\mathrm{T}}\boldsymbol{\beta}$. If each $g_{j}$ ($j=0,\ldots,p$) is constant, then the model contains only linear effects of $(\boldsymbol{X},\boldsymbol{Z})$. If $g_{j}$ is a linear function, then the model contains the linear effect of $X_j$ and the interaction effect of $\boldsymbol{U}^{\mathrm{T}}\boldsymbol{\beta}$ and $X_{j}$. The proposed model (\ref{eq:model}) accommodates many different types of outcomes. For continuous or binary outcomes, we set $f$ to be a density from the exponential family. For right-censored survival outcomes, we set $f$ to be the density under the Cox proportional hazards model.

For a sample of size $n$, the observed data consist of $(Y_{i},\boldsymbol{X}_{i},\boldsymbol{U}_{i},\boldsymbol{Z}_{i})$ for $i=1,\ldots,n$. For right-censored survival outcomes, we set $Y_{i}=(\widetilde{Y}_{i},\Delta_{i})$, where $\widetilde{Y}_{i}$ is the event or censoring time, and $\Delta_{i}$ is the event indicator. For uncensored outcomes, the log-likelihood function is $\ell_{n}(\boldsymbol{\beta},\boldsymbol{\psi},\mathcal{G})=\sum_{i=1}^{n}\log f\{Y_i;\sum_{j=0}^{p}g_{j}(\boldsymbol{U}_{i}^{\mathrm{T}}\boldsymbol{\beta})X_{ij}+\boldsymbol{Z}_{i}^{\mathrm{\mathrm{T}}}\boldsymbol{\psi}\}$, where $\mathcal{G}=(g_{0},\ldots,g_{p})$. For right-censored outcomes under the Cox model, we set $\ell_{n}$ to be the log-partial-likelihood function, such that
\[
\ell_{n}(\boldsymbol{\beta},\boldsymbol{\psi},\mathcal{G})=\sum_{i=1}^{n}\Delta_{i}\Big[\sum_{j=0}^{p}g_{j}(\boldsymbol{U}_{i}^{\mathrm{T}}\boldsymbol{\beta})X_{ij}+\boldsymbol{Z}_{i}^{\mathrm{\mathrm{T}}}\boldsymbol{\psi}-\log\Big\{\sum_{h:Y_{h}\ge Y_{i}}e^{\sum_{j=0}^{p}g_{j}(\boldsymbol{U}_{h}^{\mathrm{T}}\boldsymbol{\beta})X_{hj}+\boldsymbol{Z}_{h}^{\mathrm{\mathrm{T}}}\boldsymbol{\psi}}\Big\}\Big].
\]

\subsection{Penalized sieve estimation}
Because the likelihood involves the nonparametric functions $(g_{0},\ldots,g_{p})$, maximum likelihood estimation is not feasible. We propose to approximate $g_{j}$ by B-spline functions. Let $(B_{1},\ldots,B_{d})$ be a set of B-spline functions on a pre-specified set of grid points, such that each function passes through the origin; the construction of the B-spline functions are discussed in Appendix \ref{sec:bspline}. For $j=0,\ldots,p$, we approximate $g_{j}$ by $\gamma_{j}+\sum_{k=1}^{d}\alpha_{jk}B_{k}$, where $(\gamma_{j},\alpha_{j1},\ldots,\alpha_{jd})$ are regression parameters. For right-censored outcomes, we set $\gamma_{0}=0$ for identifiability.
%Also, if $\boldsymbol{Z}$ contains $\boldsymbol{U}$, then $g_0$ and $\boldsymbol{\psi}$ cannot be jointly identified, and we may remove one component of $\boldsymbol{U}$ in $\boldsymbol{Z}$ to ensure model identifiability.

When $p$ is large, the total number of parameters may be larger than the sample size, and penalization on $\boldsymbol{\gamma}\equiv(\gamma_{0},\ldots,\gamma_{p})^{\mathrm{T}}$ and $\boldsymbol{\alpha}\equiv(\boldsymbol{\alpha}_{jk})_{j=0,\ldots,p;k=1,\ldots,d}$ could be adopted for stable estimation and variable selection. We propose to estimate the parameters by maximizing the following penalized log-likelihood function:
\[
p\ell_{n}(\boldsymbol{\beta},\boldsymbol{\psi},\boldsymbol{\gamma},\boldsymbol{\alpha})=\ell_{n}\Big\{\boldsymbol{\beta},\boldsymbol{\psi},\big(\gamma_{j}+\sum_{k=1}^{d}\alpha_{jk}B_{k}\big)_{j=0,\ldots,p}\Big\}-\sum_{j=1}^{p}\rho_{1}(\gamma_{j};\lambda_{1})-\sum_{j=1}^p\rho_{2}(\boldsymbol{\alpha}_{j};\lambda_{2}),
\]
where $\rho_1$ and $\rho_2$ are penalty functions, $\lambda_{1}$ and $\lambda_{2}$ are tuning parameters, and $\boldsymbol{\alpha}_{j}=(\alpha_{j1},\ldots,\alpha_{jd})^\mathrm{T}$ for $j=1,\ldots,p$. This formulation allows separate selection of constant and non-constant effects of $X_{j}$ by separate penalization on $\gamma_{j}$ and $\boldsymbol{\alpha}_{j}$. Let $\widehat{\boldsymbol{\beta}}$, $\widehat{\gamma}_{j}$, and $\widehat{\boldsymbol{\alpha}}_{j}$ denote the penalized estimator of $\boldsymbol{\beta}$, $\gamma_{j}$, and $\boldsymbol{\alpha}_{j}$, respectively ($j=0,\ldots,p$). For $j=1,\ldots,p$, if $\widehat{\gamma}_{j}=0$ and $\widehat{\boldsymbol{\alpha}}_{j}=\boldsymbol{0}$, then $X_{j}$ does not have an effect on the outcome in the estimated model. If only $\widehat{\boldsymbol{\alpha}}_{j}=\boldsymbol{0}$, then $X_{j}$ has a
constant effect of $\widehat{\gamma}_{j}$. If $\widehat{\boldsymbol{\alpha}}_{j}$ is non-zero, then $X_{j}$ has a non-constant effect indexed by $\boldsymbol{U}^{\mathrm{T}}\widehat{\boldsymbol{\beta}}$.

Many choices of penalty functions, such as the (group) lasso \citep{tibshirani1996regression, yuan2006model}, smoothly clipped absolute deviation (SCAD) \citep{fan2001variable, breheny2009penalized}, and minimax concave penalty (MCP) \citep{zhang2007penalized, breheny2009penalized}, are possible. 
Although these conventional choices of penalty functions for $\rho_1$ and $\rho_2$ can produce sparse
estimation of the constant and non-constant effects, they fail to take into account the fact that $\gamma_j$ and $\boldsymbol{\alpha}_j$ ($j=1,\ldots,p$) correspond to the same predictor $X_j$.
In this paper, we propose to set
$\rho_{1}(\gamma_{j};\lambda_{1})=\lambda_{1}w_{j}\vert\gamma_{j}\vert$ and $\rho_{2}(\boldsymbol{\alpha}_{j};\lambda_{2}) =\lambda_{2}w_{j}(\boldsymbol{\alpha}_{j}^\mathrm{T}\boldsymbol{K}_{j}\boldsymbol{\alpha}_{j})^{1/2}$, where $w_j$ is a weight for the $j$th predictor, and $\boldsymbol{K}_{j}$ is some $(d\times d)$-symmetric matrix;
the first penalty is similar to the adaptive lasso penalty \citep{zou2006adaptive}, and the second penalty is a weighted version of the group lasso.
The weight $w_j$ is introduced to capture the overall signal strength
of $g_j$ and unify the degree of shrinkage of $\gamma_{j}$ and $\boldsymbol{\alpha}_{j}$.
In particular, we set $\omega_{j}=(\widetilde{\gamma}_{j}^2+\Vert\widetilde{\boldsymbol{\alpha}}_{j}\Vert^2)^{-1/2}$, where $\widetilde{\gamma}_{j}$ and $\widetilde{\boldsymbol{\alpha}}_{j}$ are estimates of $\gamma_{j}$ and $\boldsymbol{\alpha}_{j}$ obtained from maximizing the penalized log-likelihood with $w_{j}=1$ for $j=1,\ldots,p$. If the initial estimates $\widetilde{\gamma}_{j}$ and $\widetilde{\boldsymbol{\alpha}}_{j}$ are accurate in that variables with stronger signal receive smaller weights,
then the weighted estimators would yield better variable selection and estimation accuracy than unweighted estimators.

%Although the lasso is computationally efficient, the shrinkage introduced by lasso may result in substantial bias for large regression coefficients. Common nonconvex penalties such as SCAD and MCP are designed to diminish this bias while achieving selection consistency. However, these non-convex penalties tend to possess multiple local optima. We propose an adaptive-lasso type penalty \citep{zou2006adaptive} to obtain a weighted convex penalty function that corrects the intrinsic estimation bias of lasso. 

We propose to compute the estimates using an alternating algorithm. In particular, 
we initialize $\boldsymbol{\beta}$ as some unit vector
and update the parameter estimates of $(\boldsymbol{\gamma},\boldsymbol{\alpha},\boldsymbol{\psi})$ and $\boldsymbol{\beta}$ alternatively as follows until convergence.
For fixed $\boldsymbol{\beta}$, the objective function is essentially the penalized log-likelihood function for a conventional regression model under a group lasso penalty, and $(\boldsymbol{\gamma},\boldsymbol{\alpha},\boldsymbol{\psi})$ can be updated using existing algorithms for the group lasso \citep{breheny2009penalized}. For fixed $(\boldsymbol{\gamma},\boldsymbol{\alpha},\boldsymbol{\psi})$, using the Lagrange multiplier method, the (penalized) log-likelihood function is maximized at $\boldsymbol{\beta}$ such that $\partial\ell_{n}\big\{\boldsymbol{\beta},\boldsymbol{\psi},\big(\gamma_{j}+\sum_{k=1}^{d}\alpha_{jk}B_{k}\big)_{j=0,\ldots,p}\big\}/\partial\boldsymbol{\beta}+c\boldsymbol{\beta}=\boldsymbol{0}$ and $\Vert\boldsymbol{\beta}\Vert^{2}-1=0$ for some $c$. We solve for $\boldsymbol{\beta}$ and $c$ simultaneously using the Newton-Raphson algorithm.

We propose to select the tuning parameters $\lambda_{1}$ and $\lambda_{2}$ using a version of the Bayesian information criterion (BIC), defined as % \citep{schwarz1978estimating}
\[
-2\ell_{n}(\widehat{\boldsymbol{\beta}},\widehat{\boldsymbol{\psi}},\widehat{\mathcal{G}})+q\log(n^{*}),
\]
where $\widehat{\mathcal{G}}=(\widehat{\gamma}_{j}+\sum_{k=1}^d\widehat{\alpha}_{jk}B_{k})_{j=0,\ldots,p}$, $q$ is the effective degrees of freedom, and $n^{*}$ is the effective sample size. Specifically, $n^{*}=n$ for uncensored outcomes, and $n^{*}$ is the number of uncensored observations for right-censored outcomes. Following \cite{breheny2009penalized}, we define the effective degrees of freedom as
\[
q=\sum_{j=1}^{p}\bigg(\frac{\widehat{\gamma}_{j}}{\widehat{\gamma}_{j}^{*}}+\sum_{k=1}^{d}\frac{\widehat{\alpha}_{jk}}{\widehat{\alpha}_{jk}^{*}}\bigg),
\]
where $(\widehat{\gamma}_{j},\widehat{\alpha}_{jk})$ denote the estimated value of $(\gamma_{j},\alpha_{jk})$, $\widehat{\gamma}_{j}^{*}$ denote the maximizer of the unpenalized log-likelihood function with respect to $\gamma_{j}$ with other parameters fixed at the estimated value, and $\widehat{\alpha}_{jk}^{*}$ denote the maximizer of the unpenalized log-likelihood function with respect to $\alpha_{jk}$ with other parameters fixed at the estimated value. We select $(\lambda_{1},\lambda_{2})$ that yield the minimum modified BIC value.

In conventional group lasso problems, the predictor matrix of the $j$th group, denoted by $\boldsymbol{W}_{j}$, is typically transformed such that $\boldsymbol{W}_{j}^\mathrm{T}\boldsymbol{W}_{j}$ is a diagonal matrix with equal diagonal elements. This is equivalent to setting $\boldsymbol{K}_{j}$ to be (a scaled version of) $\boldsymbol{W}_{j}^\mathrm{T}\boldsymbol{W}_{j}$. In the current problem, however, the ``predictor matrix,'' which consists of rows $(X_{ij}, B_{1}(\boldsymbol{U}_{i}^\mathrm{T}\boldsymbol{\beta})X_{ij},\ldots,B_{d}(\boldsymbol{U}_{i}^\mathrm{T}\boldsymbol{\beta})X_{ij})$ ($i=1,\ldots,n$), depends on the unknown parameter $\boldsymbol{\beta}$. One estimation strategy is to set $\boldsymbol{K}_{j}$ based on the predictor matrix evaluated at some initial estimator of $\boldsymbol{\beta}$, such as that obtained under $\boldsymbol{K}_{j}=\boldsymbol{I}$. Another strategy is to update $\boldsymbol{K}_{j}$ with $\boldsymbol{\beta}$ after each iteration; this can be thought of as setting $\boldsymbol{K}_{j}$ based on the converged value of $\boldsymbol{\beta}$. Another difficulty that arises from the unknown $\boldsymbol{\beta}$ is that the converged estimates may vary with the initial value of $\boldsymbol{\beta}$. We propose to consider multiple initial values and select the final estimates that yield the smallest modified BIC.
In the simulation studies, we considered 5 initial values of $\boldsymbol{\beta}$ and updated $\boldsymbol{K}$ along with $\boldsymbol{\beta}$ at each iteration, and the algorithm converged at almost all replicates.

\section{Simulation studies}
\label{sec:sim}

\begin{figure}
\centering
\includegraphics[scale=0.85]{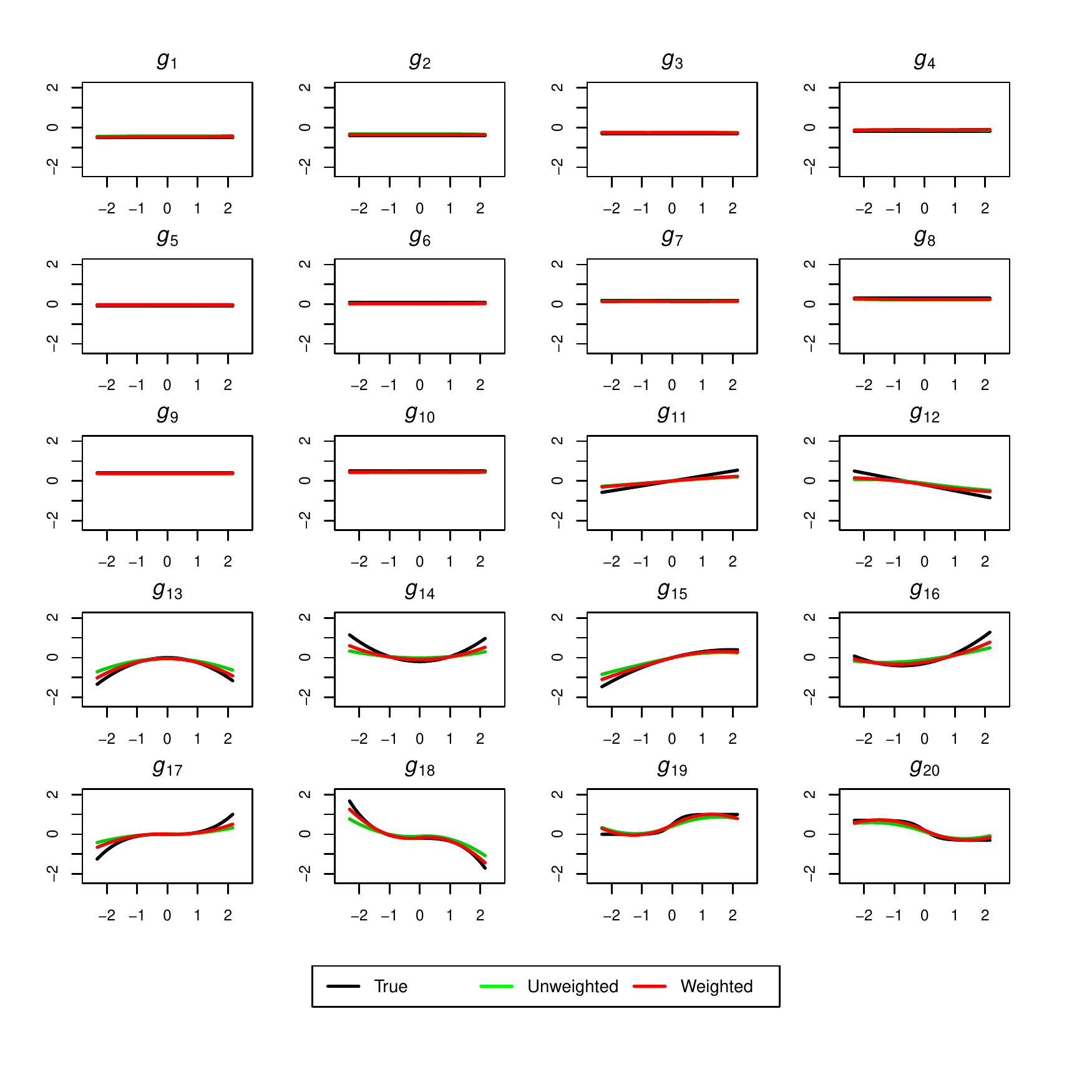}
\caption{\label{fig:sim_gaussian_p100}Estimated coefficients for the continuous outcome under $p=100$.}
\end{figure}

We set the dimension of $\boldsymbol{U}$ to be 4 and generated components of $\boldsymbol{U}$ as i.i.d. standard normal variables. We set $\boldsymbol{Z}=\boldsymbol{U}$ and generated $\boldsymbol{X}$ from the $p$-variate standard normal distribution. We set $\boldsymbol{\beta}=(0.4,-0.4,0.2,-0.8)^{\mathrm{T}}$, $\boldsymbol{\psi}=(0.2,-0.2,0.5,-0.5)^{\mathrm{T}}$, and $g_{1},\ldots,g_{20}$ to be non-zero constant, linear, or non-linear functions; the functions are plotted in Figure \ref{fig:sim_gaussian_p100}. We set $g_{0}$ and $g_{21},\ldots,g_{p}$ to be constant at 0. We considered a continuous outcome variable and a right-censored outcome variable. For the continuous outcome, we set $f(y;\mu)=(2\pi)^{-1/2}\exp\{-\frac{1}{2}(y-\mu)^2\}$, so that conditional on $(\boldsymbol{X},\boldsymbol{Z},\boldsymbol{U})$, $Y$ follows the normal distribution with unit variance. For the right-censored outcome, we set
$
f(y;\mu)=h(y)e^\mu\exp\big\{-e^\mu\int_0^yh(t)\,\mathrm{d}t\big\}$,
where $h$ is the baseline hazard function with $h(t)=t$. The censoring time was generated from an exponential distribution with the mean chosen to yield a censoring rate of about 30\%. In each setting, we considered a sample size of 500 and $p=20$, 50, and 100.

We compare the proposed methods with conventional regression models with or without interaction terms. For the proposed methods, we set the degree of the B-spline functions to be 2 and the knots at $-\max_{i}\Vert\boldsymbol{U}_{i}\Vert_{2}$, 0, and $\max_{i}\Vert\boldsymbol{U}_{i}\Vert_{2}$. We considered the proposed weighted approach and an unweighted approach with $w_{j}=1$ $(j=1,\ldots,p)$. We also considered the lasso regression on the linear predictors $(\boldsymbol{X},\boldsymbol{Z})$ and the lasso regression on $\boldsymbol{X}$, $\boldsymbol{Z}$, and pairwise interactions between components of $\boldsymbol{X}$ and $\boldsymbol{Z}$; in both cases, coefficients of $\boldsymbol{Z}$ were not penalized. In addition, we considered adaptive lasso for the models with or without interactions, where the weights are the inverse of the absolute value of the corresponding lasso estimates. In all methods, the tuning parameters were selected using the modified BIC.

We evaluate the performance of each method in terms of variable selection and prediction. For variable selection, we report the sensitivity and the false discovery rate (FDR). Sensitivity is the proportion of correctly identified signal variables among all true signal variables. FDR is the proportion of noise variables that are incorrectly identified as signal variables among all selected variables. For the proposed methods, a variable $X_{j}$ is selected if either $\gamma_{j}$ or $\boldsymbol{\alpha}_{j}$ is estimated as non-zero $(j=1,\ldots,p)$. For the proposed methods and lasso with interactions, we also report the sensitivity and FDR with respect to the selection of non-constant effects, where for the proposed methods, the non-constant effect of $X_{j}$ is selected if $\widehat{\boldsymbol{\alpha}}_{j}\ne\boldsymbol{0}$, and for lasso
with interactions, the non-constant effect is selected if the coefficient of the product of $X_{j}$ and any component of $\boldsymbol{Z}$ is non-zero.
In addition, we report the total numbers of the selected variables and the number of variables identified to have non-constant effects.

For prediction, we report the mean-squared error (MSE), defined as $\mathrm{E}(\widehat{\eta}-\eta_{0})^2$, where $\eta_{0}=\eta(\boldsymbol{\beta}_{0},\mathcal{G}_{0},\boldsymbol{\psi}_{0})$, $\eta(\boldsymbol{\beta},\mathcal{G},\boldsymbol{\psi})
\equiv\sum_{j=1}^{p}g_{j}(\boldsymbol{U}^{\mathrm{T}}\boldsymbol{\beta})X_{j}+\boldsymbol{Z}^{\mathrm{T}}\boldsymbol{\psi}$, and $(\boldsymbol{\beta}_{0},\mathcal{G}_{0},\boldsymbol{\psi}_{0})$ denote the true parameter values. For the proposed methods, $\widehat{\eta}=\eta(\widehat{\boldsymbol{\beta}},\widehat{\mathcal{G}},\widehat{\boldsymbol{\psi}})$, where $(\widehat{\boldsymbol{\beta}},\widehat{\mathcal{G}},\widehat{\boldsymbol{\psi}})$ denote the estimated parameter values. For lasso with and without interaction effects, $\widehat{\eta}=\sum_{j}\widehat{b}_{j}X_{j}+\sum_{k}\widehat{c}_{k}Z_{k}+\sum_{j,k}\widehat{d}_{jk}X_{j}Z_{k}$ and $\widehat{\eta}=\sum_{j} \widetilde{b}_{j}X_{j}+\sum_{k}\widetilde{c}_{k}Z_{k}$, respectively, where $\widehat{b}_{j}$, $\widehat{c}_{k}$, $\widehat{d}_{jk}$, $\widetilde{b}_{j}$, and $\widetilde{c}_{k}$ are the corresponding estimated regression parameters. For the right-censored outcome, we also compute the concordance index (C-index) \citep{harrell1982evaluating}, defined as $\mathrm{P}(\eta_{i}>\eta_{j}\mid\widetilde{Y}_{i}<\widetilde{Y}_{j})$ for two generic independent subjects indexed by $i$ and $j$. C-index typically takes values between 0.5 and 1, where a value of 0.5 indicates no discrimination and a value of 1 indicates perfect discrimination. For the proposed methods, we also report the absolute inner product $\vert\boldsymbol{\beta}^{\mathrm{T}}\widehat{\boldsymbol{\beta}}\vert$
to assess the estimation accuracy of $\widehat{\boldsymbol{\beta}}$. The simulation results for the continuous and right-censored outcomes based on 100 replicates are summarized in Tables \ref{Tab:sim_gaussian_lasso} and \ref{Tab:sim_cox_lasso}, respectively. Figure \ref{fig:sim_gaussian_p100} shows the average estimated values of $g_{1},\ldots,g_{20}$ for the continuous outcome under $p=100$. The simulation results under other settings are plotted in Figures \ref{fig:sim_gaussian_p50}--\ref{fig:sim_cox_p100}. 

\begin{table}
\small\centering
\caption{Simulation results for the continuous outcome.}
\begin{threeparttable}
\setlength{\tabcolsep}{1mm}
\renewcommand{\arraystretch}{1.2}
\begin{tabular}{llccccccccc} \hline
&&\multicolumn{3}{c}{$p=20$}&\multicolumn{3}{c}{$p=50$}&\multicolumn{3}{c}{$p=100$}\\\cline{3-11}
&& Proposed & Main & Interaction & Proposed & Main & Interaction & Proposed & Main & Interaction\\\hline
\multicolumn{11}{l}{Unweighted}\\\hline
SEN & Overall & 0.990 & 0.800 & 0.978 & 0.978 & 0.776 & 0.966 & 0.962 & 0.754 & 0.960\\
& Non-constant & 0.999 & - & 0.949 & 0.987 & - & 0.917 & 0.983 & - & 0.912\\
FDR & Overall & 0 & 0 & 0 & 0.377 & 0.295 & 0.504 & 0.543 & 0.479 & 0.706\\
& Non-constant & 0.253 & - & 0.426 & 0.436 & - & 0.718 & 0.530 & - & 0.833\\
NS & Overall & 19.80 & 16.01 & 19.57 & 31.57 & 22.21 & 39.13 & 42.44 & 29.30 & 65.60\\
%& Constant & 17.58 & 16.01 & 16.17 & 24.20 & 22.21 & 21.01 & 30.90 & 29.30 & 26.76\\
& Non-constant & 13.48 & 0 & 16.65 & 17.70 & 0 & 32.84 & 21.17 & 0 & 54.84\\
\multicolumn{2}{l}{$\vert\boldsymbol{\beta}^\mathrm{T}\widehat{\boldsymbol{\beta}}\vert$} & 0.997 & - & - & 0.996 & - & - &0.996 & - & -\\
\multicolumn{2}{l}{MSE} & 0.378 & 1.583 & 0.990 & 0.569 & 1.754 & 1.160 & 0.698 & 1.764 & 1.206\\\hline
\multicolumn{11}{l}{Weighted}\\\hline
SEN & Overall & 0.948 & 0.657 & 0.908 & 0.915 & 0.649 & 0.900 & 0.898 & 0.638 & 0.897\\
& Non-constant & 0.979 & - & 0.848 & 0.951 & - & 0.826 & 0.915 & - & 0.834\\
FDR & Overall & 0 & 0 & 0 & 0.155 & 0.096 & 0.324 & 0.247 & 0.200 & 0.540\\
& Non-constant & 0.069 & - & 0.243 & 0.154 & - & 0.552 & 0.222 & - & 0.709\\
NS & Overall & 18.96 & 13.14 & 18.15 & 21.79 & 14.47 & 26.90 & 24.11 & 16.19 & 39.43\\
%& Constant & 18.33 & 13.14 & 14.07 & 20.65 & 14.47 & 15.62 & 22.87 & 16.19 & 18.11\\
& Non-constant & 10.57 & 0 & 11.36 & 11.33 & 0 & 18.83 & 11.93 & 0 & 29.19\\
\multicolumn{2}{l}{$\vert\boldsymbol{\beta}^\mathrm{T}\widehat{\boldsymbol{\beta}}\vert$} & 0.998 & - & - & 0.997 & - & - &0.997 & - & -\\
\multicolumn{2}{l}{MSE} & 0.313 & 1.602 & 1.008 & 0.408 & 1.724 & 1.135 & 0.544 & 1.730 & 1.198\\\hline
\end{tabular}
\end{threeparttable}
\begin{tablenotes}
NOTE: ``SEN'' represents sensitivity; ``NS'' represents number of selected variables; ``Main'' represents lasso regression model without interactions; ``Interaction'' represents lasso regression model with interactions; ``Overall'' gives values of corresponding measures concerning all components of $\boldsymbol{X}$; ``Non-constant'' gives values of corresponding measures concerning components of $\boldsymbol{X}$ with non-constant effects on the outcome. 
\end{tablenotes}
\label{Tab:sim_gaussian_lasso}
\end{table}

\begin{table}
\small\centering
\caption{Simulation results for the right-censored outcome.}
\begin{threeparttable}
\setlength{\tabcolsep}{1mm}
\renewcommand{\arraystretch}{1.2}
\begin{tabular}{llccccccccc} \hline
& &\multicolumn{3}{c}{$p=20$}&\multicolumn{3}{c}{$p=50$}&\multicolumn{3}{c}{$p=100$}\\\cline{3-11}
& & Proposed & Main & Interaction & Proposed & Main & Interaction & Proposed & Main & Interaction\\\hline
\multicolumn{11}{l}{Unweighted}\\\hline
SEN & Overall & 0.962 & 0.795 & 0.948 & 0.909 & 0.737 & 0.914 & 0.856 & 0.689 & 0.874\\
& Non-constant & 0.916 & - & 0.895 & 0.805 & - & 0.852 & 0.702 & - & 0.794\\
FDR & Overall & 0 & 0 & 0 & 0.319 & 0.297 & 0.446 & 0.477 & 0.480 & 0.629\\
& Non-constant & 0.176 & - & 0.392 & 0.283 & - & 0.661 & 0.386 & - & 0.776\\
NS & Overall & 19.24 & 15.90 & 18.97 & 26.95 & 21.11 & 33.28 & 33.06 & 26.78 & 48.24\\
%& Constant & 17.09 & 15.90 & 15.12 & 21.92 & 21.11 & 17.97 & 26.66 & 26.78 & 20.76\\
& Non-constant & 11.28 & 0 & 14.91 & 11.42 & 0 & 25.51 & 11.84 & 0 & 36.66\\
\multicolumn{2}{l}{$\vert\boldsymbol{\beta}^\mathrm{T}\widehat{\boldsymbol{\beta}}\vert$} & 0.993 & - & - & 0.974 & - & - &0.957 & - & -\\
\multicolumn{2}{l}{MSE} & 0.843 & 1.873 & 1.419 & 1.357 & 2.142 & 1.701 & 1.518 & 2.134 & 1.746\\
\multicolumn{2}{l}{C-index} & 0.772 & 0.716 & 0.743 & 0.758 & 0.716 & 0.738 & 0.745 & 0.708 & 0.727\\\hline
\multicolumn{11}{l}{Weighted}\\\hline
SEN & Overall & 0.883 & 0.631 & 0.864 & 0.831 & 0.615 & 0.832 & 0.763 & 0.580 & 0.787\\
& Non-constant & 0.806 & - & 0.780 & 0.708 & - & 0.732 & 0.593 & - & 0.687\\
FDR & Overall & 0 & 0 & 0 & 0.136 & 0.139 & 0.303 & 0.239 & 0.273 & 0.489\\
& Non-constant & 0.073 & - & 0.238 & 0.148 & - & 0.518 & 0.217 & - & 0.658\\
NS & Overall & 17.66 & 12.62 & 17.27 & 19.35 & 14.39 & 24.13 & 20.37 & 16.19 & 31.22\\
%& Constant & 17.09 & 12.62 & 13.12 & 18.47 & 14.39 & 14.09 & 19.59 & 16.19 & 15.21\\
& Non-constant & 8.78 & 0 & 10.40 & 8.42 & 0 & 15.60 & 7.64 & 0 & 21.03\\
\multicolumn{2}{l}{$\vert\boldsymbol{\beta}^\mathrm{T}\widehat{\boldsymbol{\beta}}\vert$} & 0.994 & - & - & 0.991 & - & - & 0.982 & - & -\\
\multicolumn{2}{l}{MSE} & 0.691 & 1.822 & 1.209 & 0.929 & 1.983 & 1.414 & 1.166 & 1.947 & 1.528\\
\multicolumn{2}{l}{C-index} & 0.773 & 0.714 & 0.743 & 0.766 & 0.716 & 0.740 & 0.754 & 0.710 & 0.726\\\hline
\end{tabular}
\end{threeparttable}
\begin{tablenotes}
NOTE: See NOTE to Table \ref{Tab:sim_gaussian_lasso}.
\end{tablenotes}
\label{Tab:sim_cox_lasso}
\end{table}

In terms of prediction, both the weighted and unweighted versions of the proposed methods correctly identify the interaction structure between $\boldsymbol{X}$ and $\boldsymbol{Z}$ and yield higher prediction accuracy than other methods. In particular, they yield lower MSE in all settings and higher C-index for the right-censored outcome. In addition, the estimated value of $\boldsymbol{\beta}$ is close to the true value, indicating that the proposed methods can correctly identify the composition of the index. The weighted estimators are generally accurate, whereas the unweighted estimators tend to be biased towards zero due to the uniform shrinkage imposed on all parameters. Lasso with interaction terms generally yields smaller MSE than lasso with main effects alone, suggesting that a varying-coefficient model can be approximated by a conventional regression model with pairwise interaction terms. Nevertheless, possibly due to the complexity of the interaction model, the performance of lasso with interaction is substantially worse than that of the proposed methods.

In terms of variable selection, both the proposed methods and lasso with interactions have substantially higher sensitivity than lasso with main effects alone. The FDR is lower under the proposed methods than lasso with interactions, indicating that the proposed methods tend to yield more interpretable models. The FDR for the proposed methods are higher than those for lasso with main effects alone under some settings, possibly because lasso with main effects alone generally selects much fewer variables. For all methods, the weighted estimators yield substantially lower FDR than the unweighted estimators. By setting higher penalty for noise variables and lower penalty for signal variables, the weighted method yields higher variable selection accuracy.

%Note that the biases of the estimates for the coefficient functions can be large near boundaries due to the lack of effective sample size near the tails. Nevertheless, the overall prediction accuracy of the proposed method is still superior to alternatives that we have considered in terms of MSE and C-index. 

\section{Real data analysis}
\label{sec:rda}
\subsection{TCGA NSCLC data}
We demonstrate the application of the proposed methods using a set of NSCLC patients from TCGA. The data set consists of two subtypes of lung cancer, namely lung adenocarcinoma (LUAD) and lung squamous cell carcinoma (LUSC). We are interested in the potential risk factors associated with pulmonary function, measured by the percentage of expiratory volume in one second (FEV1); a higher FEV1 represents larger lung capacity, and patients with severely impaired lung function have an increased risk of mortality \citep{hole1996impaired}. In particular, we investigated the effects of gene expressions and clinical variables on FEV1, allowing for interactions between the two types of variables. We fit the proposed model with $\boldsymbol{U}$ consisting of age, number of packs of cigarette smoked per year or pack-year smoked (PYS), cancer subtype, tumor stage, and gender; tumor stage is dichotomized into stage I versus stage II or above. This formulation allows the effects of genomic factors to be modified by clinical variables. We set $\boldsymbol{Z} = \boldsymbol{U}$ to allow linear effects of clinical variables on FEV1. After discarding genes with zero expressions for 30\% or more subjects, the data set consists of 17 148 gene expressions. We set $\boldsymbol{X}$ to consist of 300 gene expressions that have the most significant marginal association with FEV1 (adjusted for clinical variables). After removing subjects with missing data, the sample size is 353, with 185 and 168 LUAD and LUSC patients, respectively. Following the simulation studies, we set the degree of the B-spline
functions to be 2 and the knots at $-\max_{i}\Vert\boldsymbol{U}_{i}\Vert_{2}$, 0, and $\max_{i}\Vert\boldsymbol{U}_{i}\Vert_{2}$. We adopted the weighted penalty approach. We standardized all variables to have zero mean and unit variance.

\begin{figure}
\centering
\includegraphics[scale=0.75]{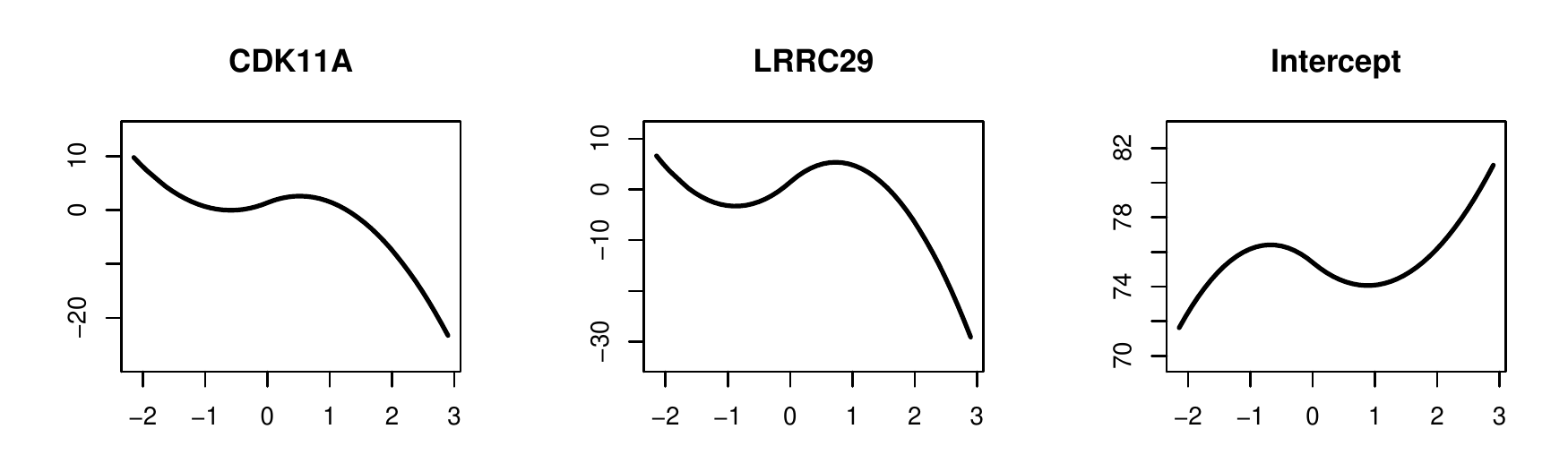}
\caption{\label{fig:rda_nsclc_lasso}Estimated coefficients for NSCLC analysis.}
\end{figure}

We identified 17 gene expressions to be associated with FEV1. The selected gene expressions and their estimated coefficients are shown in Table \ref{Tab:rda_nsclc_sel}. Among the selected gene expressions, EIF4A3 was known to be involved in the development of NSCLC, and KCNK2 and N4BP1 were known as prognostic factors in some cancer types \citep{lin2018systematic, innamaa2013expression, xu2017mir, li2019kcnk}. The effects of CDK11A and LRRC29 were identified to vary with the clinical variables; CDK11A has previously been shown to be associated with many cancer types \citep{zhou2016emerging}. The estimated index parameters $\boldsymbol{\beta}$ for age, PYS, gender, tumor stage, and cancer subtype are 0.199, 0.637, 0.157, $-$0.548, and $-$0.479, respectively. The index is dominated by PYS, tumor stage, and cancer subtype, suggesting that the effects of CDK11A and LRRC29 mainly depend on these three clinical factors. Figure \ref{fig:rda_nsclc_lasso} displays the estimated values of $g_{0}$ and the $g$ functions for CDK11A and LRRC29. 

To compare the performance of the proposed methods with existing methods, we performed a cross-validation analysis. Specifically, we repeatedly split the data set into pairs of training and validation sets 100 times, with a ratio of sample sizes of 7$:$3 and balanced clinical characteristics within each pair. In each split, we estimated the parameters using the training set and predicted the response on the validation set using the proposed methods and lasso. For the proposed methods, due to the small sample size, we fixed $\boldsymbol{\beta}$ at the estimate from the full data. For lasso, we considered regression on linear predictors ($\boldsymbol{X}$,$\boldsymbol{Z}$) and regression on $\boldsymbol{X}$, $\boldsymbol{Z}$, and pairwise interactions between components of $\boldsymbol{X}$ and $\boldsymbol{Z}$; in both cases, coefficients of $\boldsymbol{Z}$ were not penalized. The average mean-squared prediction errors over the 100 validation sets for the proposed methods, lasso without interactions, and lasso with interactions are 503.090, 512.011, and 624.602, respectively. 
% Unweighted: 481.441, 488.574, and 542.851.
This indicates that the proposed methods achieve better prediction performance over the alternatives. Between the two methods that allow for interaction effects, lasso with interactions yields much larger error than the proposed methods, probably because there are too many pairwise interaction terms to be estimated. Another possible reason is that the interaction effects cannot be adequately captured by pairwise product terms.

\subsection{TCGA LGG data}
We also applied the proposed methods to identify potential risk factors associated with the survival of patients diagnosed with lower-grade glioma (LGG) in TCGA. The data set consists of grade II and grade III tumors. Instead of integrating clinical and a single type of genomic variables, we investigated the effects of protein expressions, gene expressions, and clinical variables on time to death since initial diagnosis, allowing for interactions between protein and gene expressions. After discarding genes with zero expressions for 30\% or more subjects, the data set consists of 17 238 gene expressions. We set the overall survival time to be the outcome of interest, which is potentially right-censored. We reduced the dimension of gene expressions using principal component analysis and set $\boldsymbol{U}$ to be the first 7 principal components, which account for over 50\% of the total variability. The set of linear predictors $\boldsymbol{Z}$ consists of $\boldsymbol{U}$, age, histological grade, and gender. The set of predictors $\boldsymbol{X}$ includes the expressions of 209 proteins or phospho-proteins. After removing subjects with missing data, the sample size is 423. The median time to censoring or death is 630 days, and the censoring rate is 76.83\%.

\begin{figure}
\centering
\includegraphics[scale=0.75]{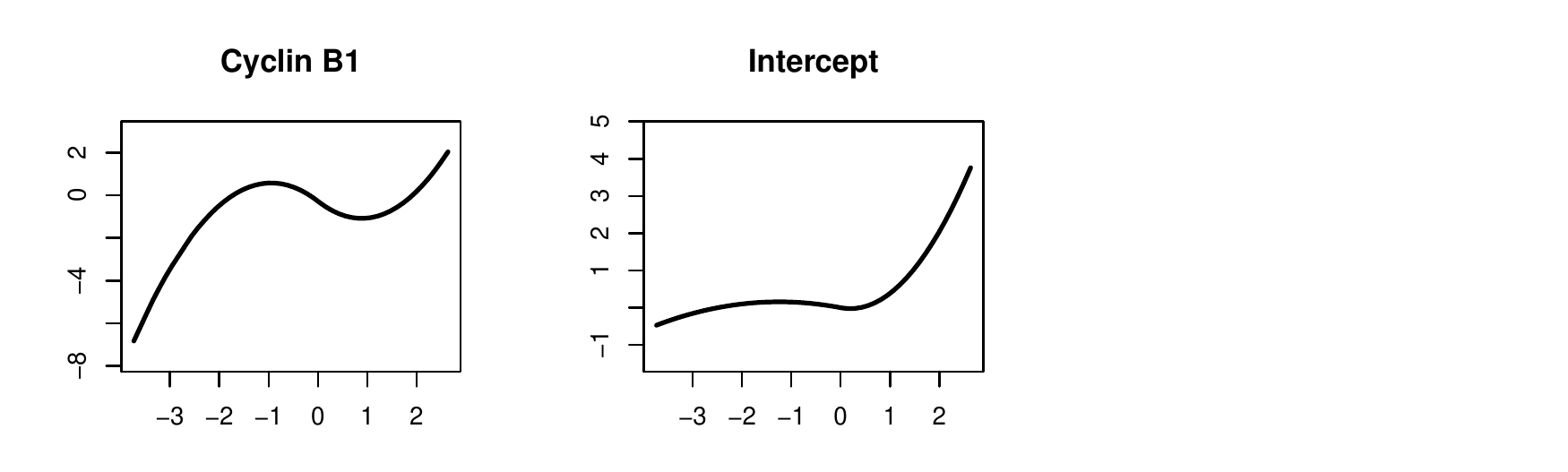}
\caption{\label{fig:rda_lgg_lasso_pc7}Estimated coefficients for LGG analysis.}
\end{figure}

We identified 7 important protein expressions to be associated with the overall survival. The selected protein expressions and their estimated coefficients are shown in Table \ref{Tab:rda_lgg_sel}. Some of the selected proteins, including FoxM1, HSP70, and Cyclin B1, have previously been shown to be associated with survival of glioma patients \citep{zhang2017akt, beaman2014reliability, chen2008overexpression}. The effect of Cyclin B1 was identified to vary with the gene expressions. 
%The estimated index parameters $\boldsymbol{\beta}$ correspond to the first 7 principal components are 0.192, 0.224, $-$0.320, 0.059, 0.343, 0.114, and 0.823 respectively. 
Figure \ref{fig:rda_lgg_lasso_pc7} displays the estimated values of $g_0$ and the $g$ function for Cyclin B1. 

Similar to the cross-validation analysis for NSCLC, we compared the prediction performance of the proposed method with lasso with and without interactions. We split the data set into pairs of training and validation sets 100 times, with a ratio of sample sizes of 7$:$3 and balanced censoring proportions and clinical characteristics within each pair. Again, for the proposed method, we fixed $\boldsymbol{\beta}$ at the estimate from the full data.
For each data split and each method, we used the estimated model from the training set to obtain survival predictions on the corresponding validation set and computed the C-index.
The average C-index values over 100 splits for the proposed methods, lasso without interactions, and lasso with interactions are 0.718, 0.704, and 0.718, respectively. 
% Unweighted: 0.731, 0.720, and 0.729
The proposed methods and lasso with interactions yield similar average C-index values, and they show slight improvement over lasso without interaction. This suggests the presence of interaction effects. By allowing those effects in the model construction, we may yield additional predictive power.

\section{Discussion}
\label{sec:diss}
In this paper, we propose a single-index varying-coefficient model for the
integration of clinical and genomic variables, where the effects of genomic variables are allowed to vary with clinical variables.
The effects of genomic variables are set as
nonparametric functions of (a projection of)
the clinical variables to accommodate intrinsically different scales of measurements between clinical and genomic variables.
Unlike the existing estimation methods for varying-coefficient models, our penalized approach separately selects for predictors with constant effects and those with varying effects. Numerical studies illustrate that the proposed methods effectively distinguish zero, constant, and non-constant effects and yield accurate prediction. %Since it is known that the lasso-type penalty tends to introduce substantial bias for large coefficients, the weighted method which allows a relatively lower penalty for large coefficient can be more predictive over the unweighted method.

The proposed methods are general and can be applied with different choices of penalty functions or outcome distributions. For example, different types of penalties, such as SCAD and MCP, can be chosen for the constant or varying effects. Also, different outcome models, such as the accelerated failure time model or additive hazard model, can be adopted.
%In this paper, we approximated the varying-coefficients using low-order basis functions to preserve the estimation efficiency. In fact, basis functions with more knot points might be used as well. Proper choice of knots and degrees is essential to avoid overfitting or underfitting the data.

There are several possible directions for future research. First, we may be interested in the interaction between two types of high-dimensional predictors, in which case the predictor vector $\boldsymbol{U}$ is high-dimensional. One possible approach is to project $\boldsymbol{U}$ to a low-dimensional space prior to fitting the proposed model. For example, as in the analysis of the LGG data, the projection can be performed by principal component analysis. However, the projected features may not have simple interpretations. Another possible approach is to perform variable selection on $\boldsymbol{U}$ by introducing an extra penalty on $\boldsymbol{\beta}$ \citep{peng2011penalized, radchenko2015high}. This approach would involve substantial computational difficulty due to the introduction of an extra penalty term. Second, it is of interest to consider more than two data types.
A possibility is to introduce extra indices that correspond to the extra data types, so that the effect of a variable may be a function of multiple indices.
This approach, however, faces enormous computational challenges because it involves multivariate nonparametric functions.

%the methods can be extended to accommodate multiple data types. The current work focuses on the integration of two data types. It is of interest to integrate multiple data types and to evaluate the interactions among them. A possible extension is to incorporate multiple indices to characterize the interaction effects among features from different data types. For example, in the analysis of LGG data, we may incorporate an additional index corresponds to the clinical factors. However, it probably increases the model complexity and is more computationally intensive. 

%\section*{Declaration of competing interest}
%The authors declare that there is no conflict of interest.

%\section*{Funding}
%This research received no specific grant from any funding agency in the public, commercial, or not-for-profit sectors.

\bibliographystyle{apa}
\bibliography{ref}

\appendix
\counterwithin{figure}{section}
\counterwithin{table}{section}
\section{Construction of basis functions}
\label{sec:bspline}
We discuss the construction of 2-degree basis functions
that pass through the origin and are continuously differentiable;
basis functions of a general degree can be constructed analogously.
Let $(k_{1},\ldots,k_{d})$ be an ordered set of grid points, where the number of grid points $d$ is odd and is larger than 2, and $k_{(d+1)/2}=0$. Let $d'=(d+1)/2$, $(\widetilde{L}_1,\ldots,\widetilde{L}_{d'})$ be a set of 2-degree B-spline functions on $(0,-k_{d'-1},\ldots,-k_{1})$, and $(R_{1},\ldots,R_{d'})$ be a set of 2-degree B-spline functions on $(0,k_{d'+1},\ldots,k_{d})$. All B-spline functions do not have an intercept, such that $\widetilde{L}_{1}(0)=\cdots=\widetilde{L}_{d'}(0)=R_{1}(0)=\cdots=R_{d'}(0)=0$. Let $L_{j}=\widetilde{L}_{j}(-x)$ for $j=1,\ldots,d'$.
The set of continuously differentiable spline functions spanned by these B-spline functions is therefore
\[
\bigg\{f=\sum_{j=1}^{d'}c_jL_j+\sum_{j=1}^{d'}c_{j+d'}R_j:(c_1,\ldots,c_{2d'})\in\mathbb{R}^{2d'},\sum_{j=1}^{d'}c_jL_j^{(1)}(0)=\sum_{j=1}^{d'}c_{j+d'}R_j^{(1)}(0)\bigg\},
\]
where $h^{(1)}$ denotes the first derivative of the function $h$. We can then construct the basis function as
\[
\Big(L_{1}+\frac{k_{d'+1}}{k_{d'-1}}R_{1},L_{2},\ldots,L_{d'},R_2,\ldots,R_{d'}\Big).
\]

\section{Additional simulation results}
\begin{figure}[H]
\centering
\includegraphics[scale=0.85]{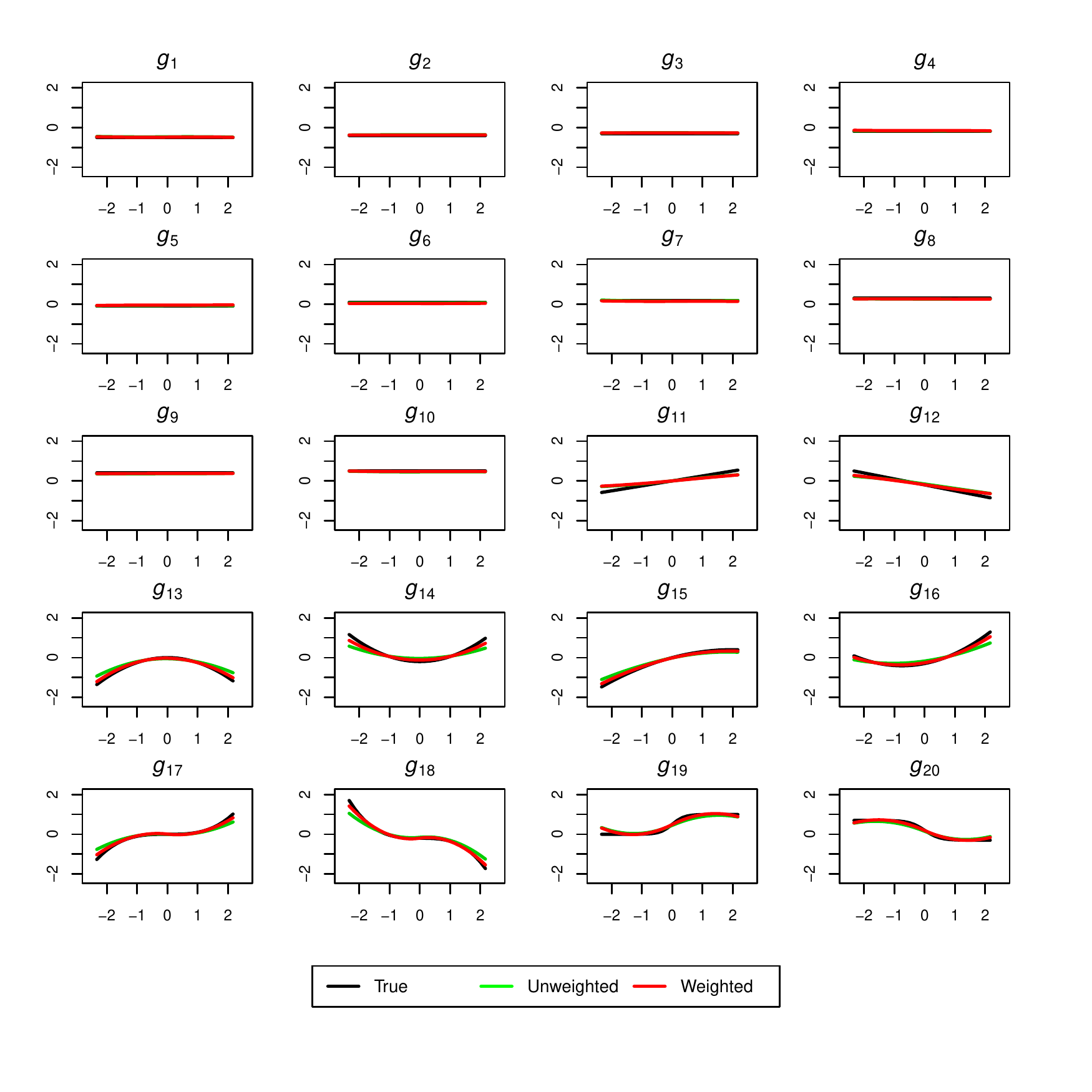}
\caption{\label{fig:sim_gaussian_p20}Estimated coefficients for the continuous outcome under $p=20$.}
\end{figure}

\begin{figure}[H]
\centering
\includegraphics[scale=0.85]{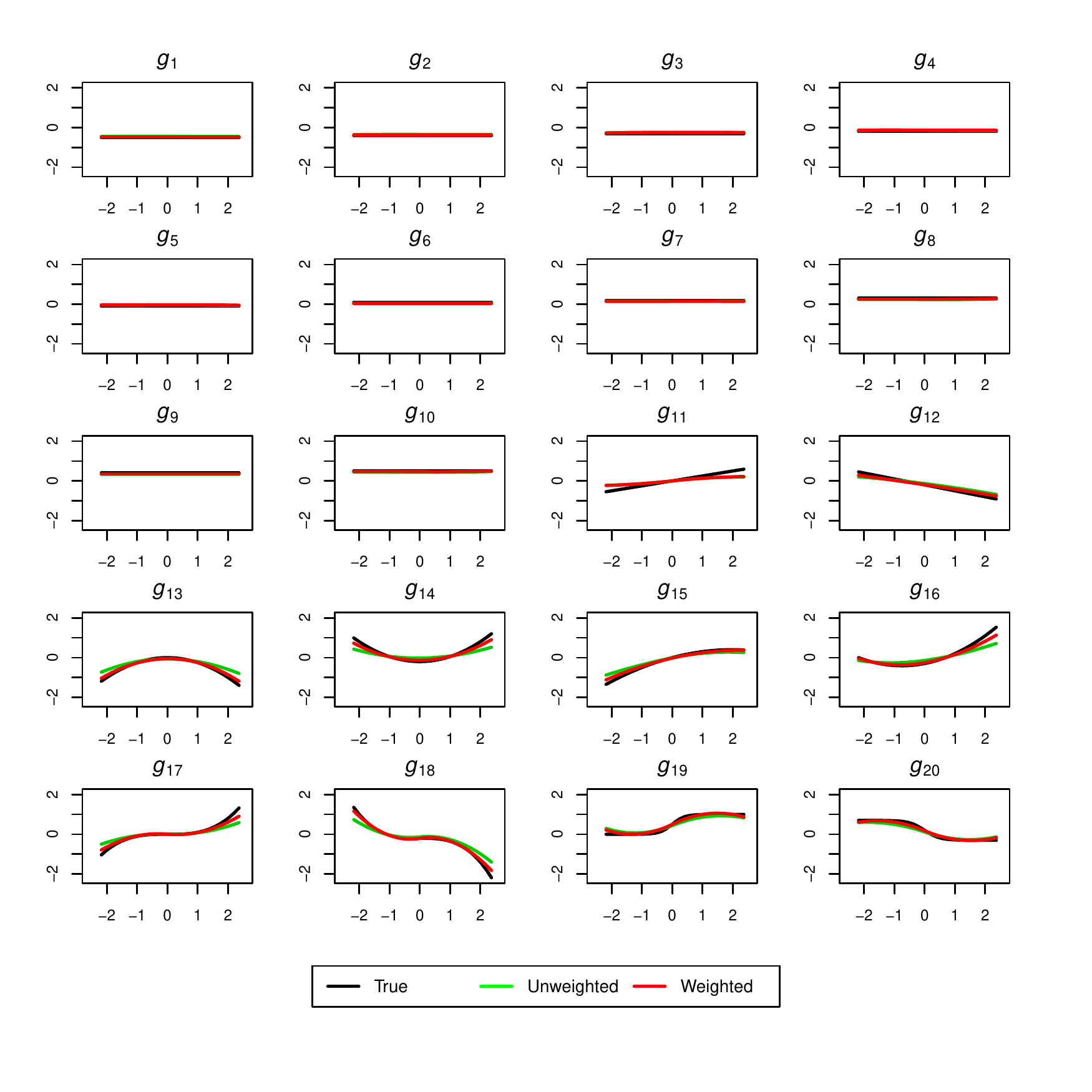}
\caption{\label{fig:sim_gaussian_p50}Estimated coefficients for the continuous outcome under $p=50$.}
\end{figure}

\begin{figure}[H]
\centering
\includegraphics[scale=0.85]{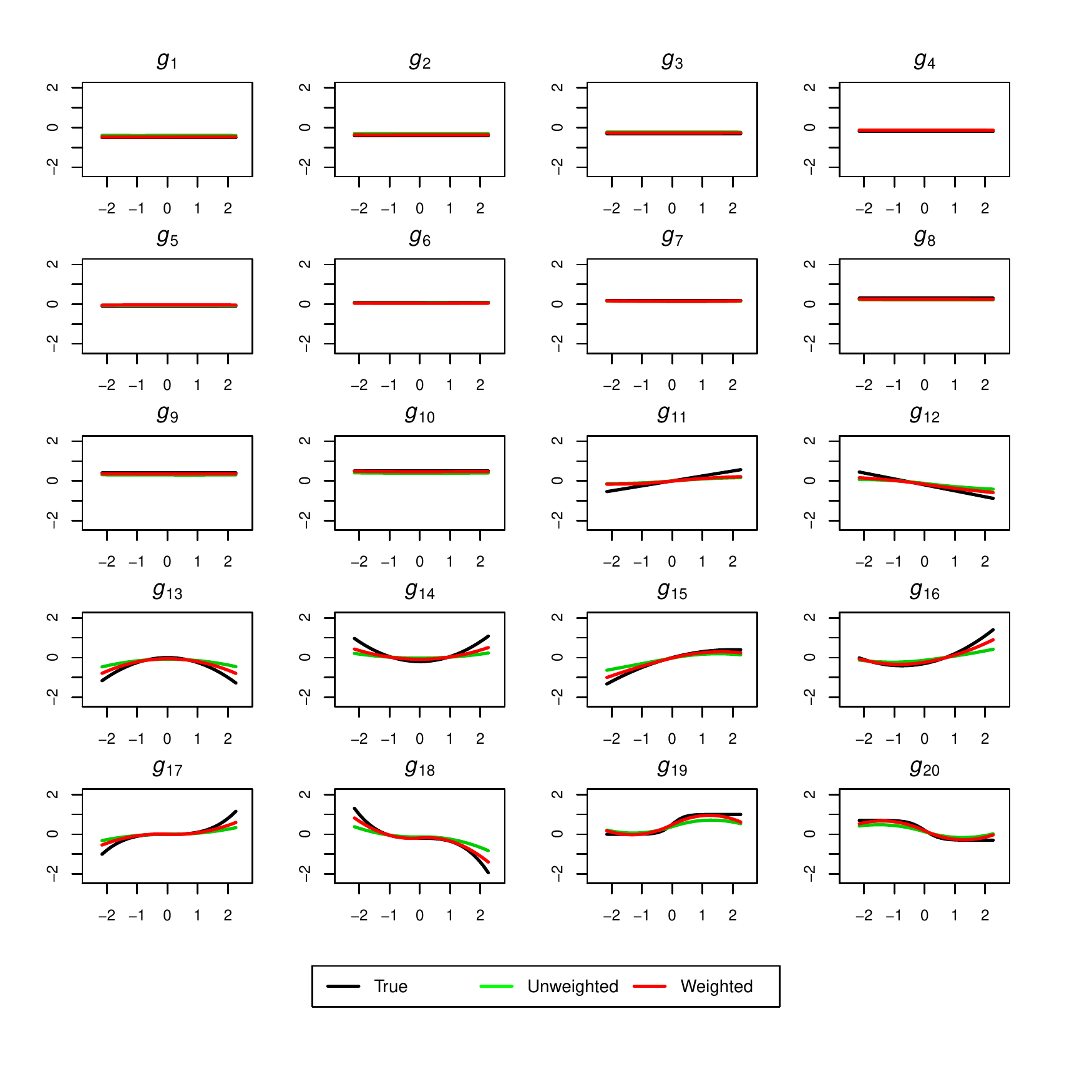}
\caption{\label{fig:sim_cox_p20}Estimated coefficients for the right-censored outcome under $p=20$.}
\end{figure}

\begin{figure}[H]
\centering
\includegraphics[scale=0.85]{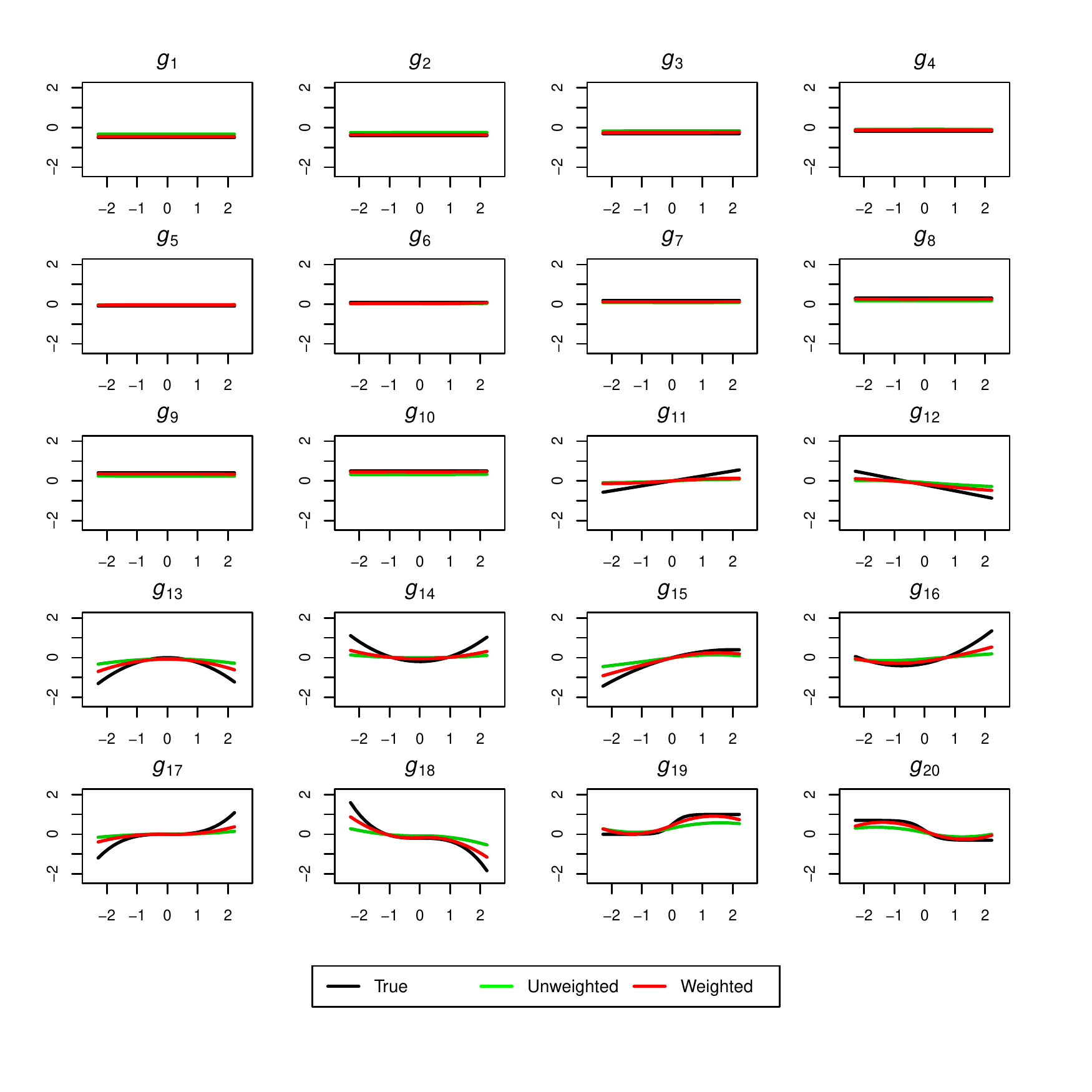}
\caption{\label{fig:sim_cox_p50}Estimated coefficients for the right-censored outcome under $p=50$.}
\end{figure}

\begin{figure}[H]
\centering
\includegraphics[scale=0.85]{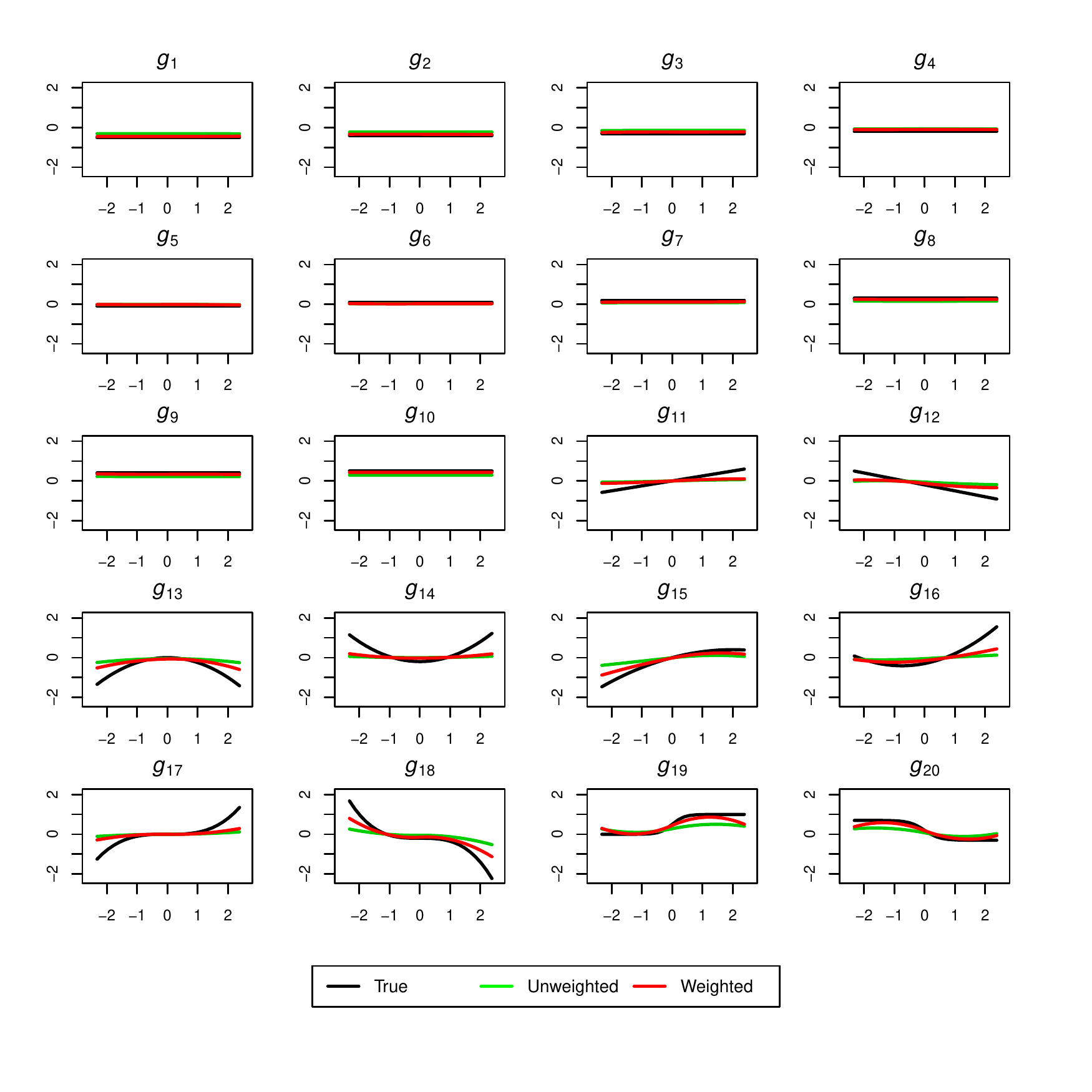}
\caption{\label{fig:sim_cox_p100}Estimated coefficients for the right-censored outcome under $p=100$.}
\end{figure}

\section{Additional real data analysis results}
\begin{table}[H]
\small\centering
\caption{Selected gene expressions for the NSCLC analysis.}
\begin{tabular}{lr} \hline
Gene & Coefficient\\\hline
ANKRD13D & 0.143\\
CDK11A & (varying)\\
CRELD2 & 1.671\\
C12orf56 & $-$1.611\\
C8orf38 & 2.008\\
C8orf58 & 2.348\\
EIF4A3 & $-$0.069\\
KCNK2 & $-$2.506\\
LOC642826 & 1.233\\
LRRC29 & (varying)\\
LRRTM2 & $-$1.168\\
NRN1L & 1.553\\
N4BP1 & 1.449\\
PLEKHG4B & 0.724\\
RNF122 & $-$3.429\\
THAP4 & 1.591\\
ZNF75D & 2.656\\\hline
\end{tabular}
\label{Tab:rda_nsclc_sel}
\end{table}

\begin{table}[H]
\small\centering
\caption{Selected protein expressions for the LGG analysis.}
\begin{tabular}{lr} \hline
Protein & Coefficient\\\hline
Cyclin B1 & (varying)\\
FoxM1 & $-$0.001\\
HER3 & $-$0.149\\
HSP70 & $-$0.393\\
MRE11 & $-$0.241\\
Stathmin & 0.251\\
ERCC5 & 0.257\\\hline
\end{tabular}
\label{Tab:rda_lgg_sel}
\end{table}

\end{document}